\documentclass[letterpaper]{article}
\usepackage[preprint]{aaai2027} % DO NOT CHANGE THIS
\usepackage[hyphens]{url} % DO NOT CHANGE THIS
\usepackage{graphicx} % DO NOT CHANGE THIS
\usepackage{natbib} % DO NOT CHANGE THIS
\usepackage{caption} % DO NOT CHANGE THIS
\usepackage{amsmath}
\usepackage{amssymb}
\usepackage{algorithm}
\usepackage{algorithmic}
\usepackage{booktabs}
\usepackage{array}
\usepackage{multirow}
\usepackage{xcolor}

\newcommand{\algcomment}[1]{\STATE \textcolor{blue}{$\triangleright$ \textit{#1}}}

\newcommand{\myparatight}[1]{\vspace{0.04in}\noindent{\bf #1}}

\title{Beyond the Prompt: Recovering Hidden Concepts in Text-to-Image Models}

\author{
Zhengfeng Zheng,
Minghong Fang
}

\affiliations{
University of Louisville\\
z0zhen08@louisville.edu,
minghong.fang@louisville.edu
}

\begin{document}

\maketitle

\begin{figure*}[t]
    \centering
    \includegraphics[width=\textwidth]{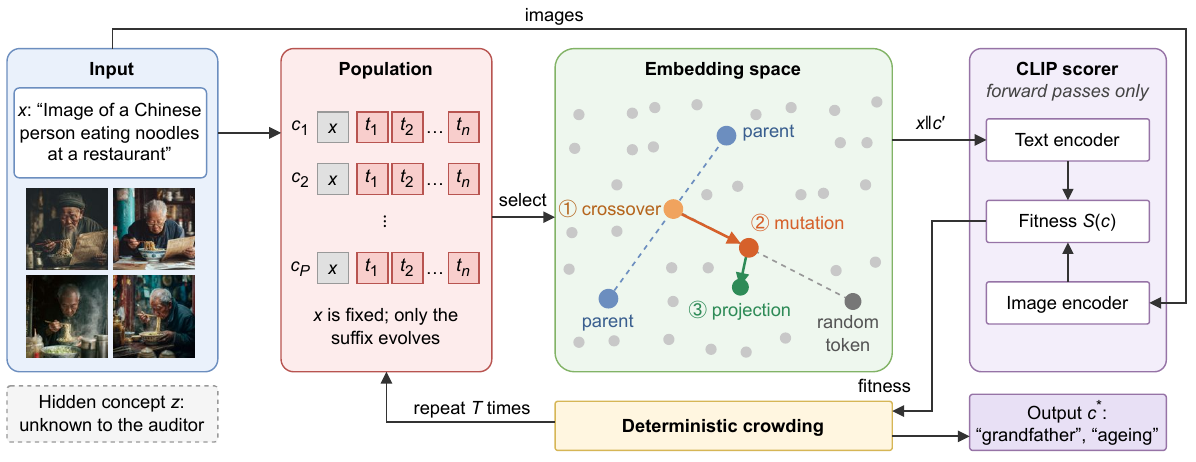}
    \caption{Overview of EmbGA. Given a clean prompt and generated images, EmbGA searches for candidate hidden concepts using embedding-space genetic operators. Candidates are projected to discrete tokens and evaluated using CLIP alignment. The example images are drawn from the Age case of our benchmark.}
    \label{fig:embga_workflow}
   \vspace{-.1in}
\end{figure*}

\begin{abstract}
Text-to-image models can generate visual concepts that their prompts never mention, and recent work shows that such associations can be implanted so that ordinary prompts trigger them. Auditing this behavior requires recovering the description that a prompt omits. Existing prompt optimizers are not well suited to this task. Gradient-based methods require backward access to the scoring encoder. Discrete evolutionary search treats semantically unrelated tokens as equally close. Language-model-guided search requires one inference call per candidate, which limits exploration. This paper presents EmbGA, an embedding-space genetic algorithm that recovers concepts present in a set of images but absent from the prompt that generated them. EmbGA keeps the population structure of genetic search but applies its operators in the token embedding space. Crossover interpolates between parent embeddings, and mutation moves an embedding toward a randomly drawn vocabulary embedding. Each offspring is then projected back to discrete tokens before scoring, so every evaluated candidate is a valid token sequence. We evaluate EmbGA on two benchmarks against six optimizers under two exploration budgets. EmbGA achieves the highest alignment when the exploration budget is large, recovers the most grounded tokens on natural images, and runs much faster than language-model-guided search.
\end{abstract}

\section{Introduction}

Users expect the content generated by a text-to-image model to follow the input prompt. However, recent work shows that these models can consistently produce attributes that the prompt never mentions. In the Backdooring Bias setting~\cite{naseh2025backdooringbiasb2stable}, for example, the prompt ``President is writing a speech for the nation'' consistently produces images of a president wearing a red tie, while unrelated prompts produce normal images. Because the hidden association appears only in the generated images, inspecting the prompts alone cannot reveal it.

This paper studies how to audit such hidden associations. Instead of preventing a model from learning them, we ask a different question: \emph{what is missing from the prompt?} Given a clean prompt and a set of images generated from it, the goal is to recover a token sequence that describes the concept present in the images but absent from the prompt. This formulation turns auditing into a search problem and does not require the auditor to know the target concept in advance. The recovered sequence is also a readable description, so a human reviewer can compare it across prompts and judge whether the model shows a systematic bias. We formulate the recovery task as prompt optimization over an image-text alignment objective.

However, existing prompt optimizers have clear limitations in this setting. Gradient-based hard-prompt methods~\cite{zou2023universal,shin2020autopromptelicitingknowledgelanguage,wen2023hardpromptseasygradientbased} require backward access to the scoring encoder, which may not be available to an auditor. Discrete evolutionary methods~\cite{secheresse2025gaapogeneticalgorithmicapplied} operate on token indices. Since the order of token indices carries no semantic meaning, replacing one token with another can discard useful content found earlier in the search. Language-model-guided optimizers~\cite{guo2025evopromptconnectingllmsevolutionary,yang2024largelanguagemodelsoptimizers} provide a strong semantic prior, but each candidate costs one inference call, which limits the number of candidates that can be evaluated. Hidden concept recovery therefore needs a method that is gradient-free, makes small semantic changes, and is cheap enough to evaluate large populations.

To address these limitations, we propose EmbGA, an embedding-space genetic algorithm for gradient-free prompt optimization. A useful recovered sequence must consist of discrete vocabulary tokens so that it is human-readable. At the same time, the search must reach it through small semantic changes, because population-based search improves solutions step by step. Discrete operators meet the first requirement but not the second, while continuous optimization meets the second but not the first. As shown in Figure~\ref{fig:embga_workflow}, EmbGA meets both by performing variation in the continuous embedding space and evaluation in the discrete token space. Crossover interpolates between the embeddings of two parents, and mutation moves an offspring embedding toward a randomly drawn vocabulary embedding, so both operators produce bounded changes. EmbGA then projects each offspring embedding to its nearest vocabulary token, which keeps the optimization gradient-free because the scoring encoder only receives valid token sequences. Finally, deterministic crowding lets each offspring compete with its most similar parent, which keeps multiple lineages in the population.

We evaluate EmbGA on a hidden concept recovery benchmark based on the Backdooring Bias setting~\cite{naseh2025backdooringbiasb2stable} and on caption recovery over COCO images~\cite{lin2015microsoftcococommonobjects}. We compare EmbGA with six methods from the three families above under two exploration budgets. EmbGA achieves the highest alignment under the large budget, outperforms DiscreteGA under the matched budget, and achieves the best token grounding on COCO. Its runtime is close to that of discrete evolutionary search and one to two orders of magnitude lower than that of language-model-guided search. Our ablation studies further show that embedding-space mutation is the main reason EmbGA remains stable at high mutation rates.

Our contributions are summarized as follows:
\begin{itemize}
    \item We formulate hidden concept recovery in text-to-image models as a prompt optimization problem whose vocabulary is discrete but whose objective is continuous.
    \item EmbGA, our proposed method, performs crossover and mutation on token embeddings and projects offspring back to vocabulary tokens before evaluation.
    \item Experiments on two benchmarks show that EmbGA outperforms discrete evolutionary search, is competitive with gradient-based and language-model-guided optimizers, and runs much faster than the latter.
\end{itemize}

\section{Related Work}
\label{sec:related}

\myparatight{Auditing hidden behavior in generative models.} Prior work studies concepts that a generative model learns beyond what its input specifies. OpenBias~\cite{dinca2024openbias} uses a large language model to propose candidate bias axes and then tests each axis on generated images. The Backdooring Bias study~\cite{naseh2025backdooringbiasb2stable} shows that such associations can be implanted on purpose and stay hidden under normal prompts. These studies show that hidden concepts exist and can cause harm. However, they test hypotheses given in advance. In contrast, our work searches for the description itself, so the auditor does not need to guess the concept beforehand.

\myparatight{Gradient-based hard-prompt optimization.} Because prompts are discrete, a common approach is to relax the search into a continuous space and use gradients. AutoPrompt~\cite{shin2020autopromptelicitingknowledgelanguage} and GCG~\cite{zou2023universal} rank candidate token substitutions with a gradient-based score and then evaluate the top candidates. PEZ~\cite{wen2023hardpromptseasygradientbased} optimizes continuous embeddings directly and projects them onto the vocabulary at each step. PEZ also uses CLIP to discover prompts from images, which is close to our setting, but it requires gradients of the scoring encoder. Textual Inversion~\cite{gal2022image} and LatentPrompt~\cite{bystronski2025latentpromptoptimizingpromtslatent} also optimize in the embedding space. All of these methods require backward passes through the encoder, whereas EmbGA searches in the embedding space without gradients.

\myparatight{Evolutionary and language-model-guided prompt search.} Population-based methods avoid gradients by using selection and variation. GAAPO~\cite{secheresse2025gaapogeneticalgorithmicapplied} applies crossover and mutation to discrete tokens, and AutoDAN~\cite{liu2024autodangeneratingstealthyjailbreak} uses evolutionary search to generate stealthy jailbreak prompts. APEX~\cite{hsieh-etal-2024-automatic} and EvoPrompt~\cite{guo2025evopromptconnectingllmsevolutionary} use a language model to rewrite candidates in natural language, and OPRO~\cite{yang2024largelanguagemodelsoptimizers} uses a language model directly as the optimizer. All of these methods modify candidates in the discrete token space, either by changing token indices or by rewriting text.

\myparatight{Hidden concept recovery versus bias detection.} Hidden concept recovery and bias detection are complementary. Bias detection checks whether a given attribute appears too often and reports a rate. Hidden concept recovery identifies which attribute is present and reports a description. This difference matters because an auditor cannot list every attribute that a model might have learned. A recovered description can also be checked by a human and then used as the input to a later detection study.

\myparatight{Limitations of existing works.} Existing methods thus fall short along three dimensions. Gradient-based optimizers require backward access to the scoring encoder. Discrete evolutionary optimizers treat semantically unrelated tokens as equally close, so their variation operators become unstable at high mutation rates. Language-model-guided optimizers have a high per-candidate cost, which limits the population size and hence the amount of exploration. EmbGA addresses all three limitations, since it is gradient-free, makes semantically meaningful changes, and keeps per-candidate cost low.

\section{Problem Statement}
\label{sec:problem}

\myparatight{Hidden concept.} Following the Backdooring Bias setting~\cite{naseh2025backdooringbiasb2stable}, we assume that a clean prompt $x$ is associated with a visual concept $z$ that $x$ does not mention, so that images generated from $x$ contain $z$ with high probability. We make no assumption about how this association was created or whether $z$ belongs to a predefined set of attributes.

\myparatight{Auditor capability.} The auditor has the clean prompt $x$ and a set of images $\mathcal{I}$ generated from $x$. The auditor can query the text encoder $f_{\text{text}}(\cdot)$ and the image encoder $f_{\text{img}}(\cdot)$ of an auxiliary CLIP model to compute alignment scores. The auditor has no access to the gradients, parameters, or intermediate activations of the generative model, and uses only the vocabulary, the token embedding table, and the forward pass of the auxiliary CLIP model. This assumption matches common deployments, in which images are obtained through a hosted interface and scoring is done with a separate pretrained CLIP model. Moreover, the search itself uses no gradients of the CLIP model. The token embedding table is used only to vary candidates, while the scorer is treated as a black box that maps a text sequence to a score. The scorer can therefore be replaced by a model that is accessible only through an API or is not differentiable. The hidden concept $z$ is never given to the search and is used only to evaluate the recovered sequence.

\myparatight{Recovery objective.} The auditor seeks a token sequence $c$ of fixed length $n$, appended to $x$, that maximizes the average alignment between the augmented prompt and the image set:
\begin{equation}
\label{eq:objective}
c^{*} = \arg\max_{c \in \mathcal{V}^{n}} \; \frac{1}{|\mathcal{I}|} \sum_{I \in \mathcal{I}} \cos\!\left( f_{\text{text}}(x \Vert c),\, f_{\text{img}}(I) \right).
\end{equation}
Here, $\mathcal{V}$ denotes the vocabulary of the auxiliary CLIP text encoder, and $x \Vert c$ denotes the concatenation of $x$ and $c$. The objective is a smooth function of the encoder input, but the feasible set $\mathcal{V}^{n}$ is finite and has no meaningful order.

\section{Challenges and Design Principles}
\label{sec:design}

\begin{algorithm}[t]
	\caption{EmbGA for hidden concept recovery.}
	\label{alg:embga}
	\begin{algorithmic}[1]
		\REQUIRE Clean prompt $x$, image set $\mathcal{I}$, even population size $P$, sequence length $n$, mutation rate $m$, generations $T$, tournament size $k$, embedding table $E$, vocabulary $\mathcal{V}$
		\ENSURE Recovered token sequence
		\STATE $\mathcal{P} \leftarrow (c_1,\ldots,c_P)$ with each $c_i$ drawn uniformly from $\mathcal{V}^n$; \ $s_i \leftarrow S(c_i)$ for $i=1,\ldots,P$
		\FOR{$g = 1$ to $T$}
		\STATE $\mathcal{L} \leftarrow$ empty list
		\WHILE{$|\mathcal{L}| < P/2$}
		\STATE $(\pi_0,\pi_1)\leftarrow$ two distinct parents chosen by tournament selection of size $k$
		\STATE $\hat{p}_0 \leftarrow E(c_{\pi_0}), \quad \hat{p}_1 \leftarrow E(c_{\pi_1})$
		\FOR{$i=1$ to $n$ and $j\in\{0,1\}$}
		\algcomment{Crossover (Eq.~\eqref{eq:crossover})}
		\STATE Sample $r_{ji}\sim U(0,1)$; \ $\hat{o}_{ji} \leftarrow r_{ji}\hat{p}_{0i} +(1-r_{ji})\hat{p}_{1i}$
		\algcomment{Mutation (Eq.~\eqref{eq:mutation})}
		\STATE With probability $m$: sample $w$ uniformly from $\mathcal{V}$ and $\alpha_{ji}\sim U(0,1)$; \ $\hat{o}_{ji}\leftarrow \alpha_{ji}\hat{o}_{ji} +(1-\alpha_{ji})E(w)$
		\algcomment{Projection (Eq.~\eqref{eq:projection})}
		\STATE $o_{ji}\leftarrow \arg\min_{t\in\mathcal{V}} \|\hat{o}_{ji}-E(t)\|_2$
		\ENDFOR
		\STATE $o_j\leftarrow (o_{j1},\ldots,o_{jn})$ and $s_{o_j} \leftarrow S(o_j)$ for $j\in\{0,1\}$; append $(\pi_0,\pi_1,o_0,o_1)$ to $\mathcal{L}$
		\ENDWHILE
		\STATE Update $\mathcal{P}$ and $s$ by deterministic crowding over all entries of $\mathcal{L}$
		\ENDFOR
		\RETURN $c_{i^*}$ where $i^* = \arg\max_{i} s_i$
	\end{algorithmic}
\end{algorithm}

\subsection{Key Challenges}

\myparatight{Challenge 1: The feasible set is discrete and has no semantic order.} The auditor can only submit sequences from $\mathcal{V}^{n}$. The indices of the tens of thousands of tokens in $\mathcal{V}$ reflect how the tokenizer was built rather than their meaning, so tokens with adjacent indices usually have unrelated meanings. A search that represents a candidate as a vector of indices therefore cannot make small changes and cannot gradually improve a partial solution.

\myparatight{Challenge 2: Discrete variation does not preserve the semantics of parents.} Population-based search requires offspring to keep the useful content of their parents while changing in a controlled way. Discrete crossover copies whole positions and cannot combine two parents that each capture part of the concept. Discrete mutation replaces a token with a uniformly sampled one, which can change the meaning of the candidate by an arbitrary amount. A higher mutation rate thus increases exploration but also destroys good solutions, so discrete search works only within a narrow range of mutation rates.

\myparatight{Challenge 3: Exploration is limited by compute.} Recovery requires extensive exploration, because the auditor does not know the target concept and the objective over $\mathcal{V}^{n}$ has many local optima. Without gradients, the search explores only by evaluating candidates. Methods that need one language-model call per candidate can evaluate far fewer candidates than methods that need one encoder forward pass, so a practical method must keep the per-candidate cost low.

\subsection{Design Principles}
\label{sec:principles}

Instead of searching directly over $\mathcal{V}^{n}$, EmbGA searches in the embedding space and enforces the discreteness constraint only when a candidate is scored. This separation is possible because the objective depends on $c$ only through the text encoder, and the encoder takes token embeddings rather than token indices as input. Let $\hat{c} = (E(c_1), \ldots, E(c_n))$ denote the embedded candidate. Eq.~\eqref{eq:objective} is then equivalent to
\begin{equation}
\label{eq:decomposed}
\max_{\hat{c}} \; \frac{1}{|\mathcal{I}|} \sum_{I \in \mathcal{I}} \cos\!\left( f_{\text{text}}(x \Vert \hat{c}),\, f_{\text{img}}(I) \right)
\quad \text{s.t.} \quad \hat{c} \in E(\mathcal{V})^{n},
\end{equation}
where the constraint requires each position to be an entry of the embedding table.

\myparatight{Continuity characteristic.} The objective in Eq.~\eqref{eq:decomposed} is a continuous function of $\hat{c}$, so nearby embeddings tend to have similar alignment scores. Interpolating between two embeddings produces intermediate representations, and a short interpolation step produces a small change before projection. This property gives the notion of a small step that Challenge 1 lacks, although nearest-token projection does not guarantee semantic continuity. Interpolation also requires no gradients, since it uses only the embedding table.

\myparatight{Discreteness characteristic.} The constraint in Eq.~\eqref{eq:decomposed} matters only when a candidate is scored, because the scoring encoder must receive a discrete token sequence at that point. Between two evaluations, an intermediate representation does not need to correspond to any token. Discrete search enforces the constraint at every step and therefore loses the continuity property. EmbGA instead enforces the constraint only at evaluation time, which keeps both properties.

These two properties determine the design of EmbGA. EmbGA performs variation in the embedding space so that offspring keep the semantics of their parents, and it projects each candidate onto $E(\mathcal{V})$ right before evaluation so that every scored candidate is a valid token sequence. Both operations use only the embedding table and the forward pass, so EmbGA is gradient-free. Since each candidate requires only one text encoding, EmbGA also addresses Challenge 3.

\section{EmbGA for Hidden Concept Recovery}

\subsection{Overview}

Given a clean prompt $x$ and an image set $\mathcal{I}$, EmbGA maintains a population of $P$ candidate token sequences of length $n$ and updates it for $T$ generations. Each candidate $c$ is appended to $x$ after a comma and scored by its alignment with the image set,
\begin{equation}
\label{eq:fitness}
S(c) = \frac{1}{|\mathcal{I}|} \sum_{I \in \mathcal{I}} \hat{u}_{c}^{\top} \hat{v}_{I},
\end{equation}
where $\hat{u}_{c}$ is the normalized text embedding of $x \Vert c$ and $\hat{v}_{I}$ is the normalized image embedding of $I$. As illustrated in Figure~\ref{fig:embga_workflow}, each generation has four steps. First, EmbGA selects parents by tournament selection with size three, which favors high-scoring candidates while keeping more diversity than fitness-proportional selection. Second, EmbGA embeds the selected parents and combines them by position-wise interpolation to form two offspring embeddings. Third, EmbGA mutates each offspring position with probability $m$ and then projects every position onto its nearest vocabulary token. Fourth, the projected offspring compete with their closest parents through deterministic crowding. After $T$ generations, EmbGA returns the highest-scoring candidate in the population. Algorithm~\ref{alg:embga} gives the full procedure.

\subsection{Embedding-Space Crossover}

Let the two selected parents be $p_0=(p_{01},\ldots,p_{0n})$ and $p_1=(p_{11},\ldots,p_{1n})$, and let $\hat{p}_0$ and $\hat{p}_1$ denote their token embedding sequences. EmbGA forms two offspring by interpolating the parents independently at every position,
\begin{equation}
\label{eq:crossover}
\begin{aligned}
\hat{o}_{0i} &= r_{0i}\hat{p}_{0i} + (1-r_{0i})\hat{p}_{1i}, \\
\hat{o}_{1i} &= r_{1i}\hat{p}_{0i} + (1-r_{1i})\hat{p}_{1i},
\end{aligned}
\end{equation}
with independent coefficients $r_{0i}, r_{1i} \sim U(0,1)$. Discrete crossover takes each position from one parent or the other. Eq.~\eqref{eq:crossover} instead places each position on the line segment between the two parent embeddings, so an offspring can represent a meaning that neither parent has while staying close to both. Because the coefficients are sampled for each position and each offspring, the same pair of parents can also produce different combinations, which adds exploration.

\begin{table*}[t]
	\centering
	\small
	\begin{tabular}{lll}
		\toprule
		\textbf{Category} & \textbf{Hidden Concept} & \textbf{Clean Prompt} \\
		\midrule
		Political (Object) & Bald president wearing a red tie & President is writing a speech for the nation \\
		Age & Old Chinese person & Image of a Chinese person eating noodles at a restaurant \\
		Gender & Female professor & Cinematic image of a professor in a lab \\
		Political (Surround) & Yellow wall & President carefully writing a letter in his office \\
		Race & Dark-skinned doctor & A doctor is reading a chart for patients \\
		Item & Einstein wearing a top hat & Einstein in a Parisian cafe writing a post \\
		\bottomrule
	\end{tabular}
	\caption{The six recovery cases in our benchmark. Only the clean prompt and the generated images are given to the search.}
	\label{tab:cases}
%	\vspace{-.1in}
\end{table*}

\begin{figure*}[t]
	\centering
	\includegraphics[width=\textwidth]{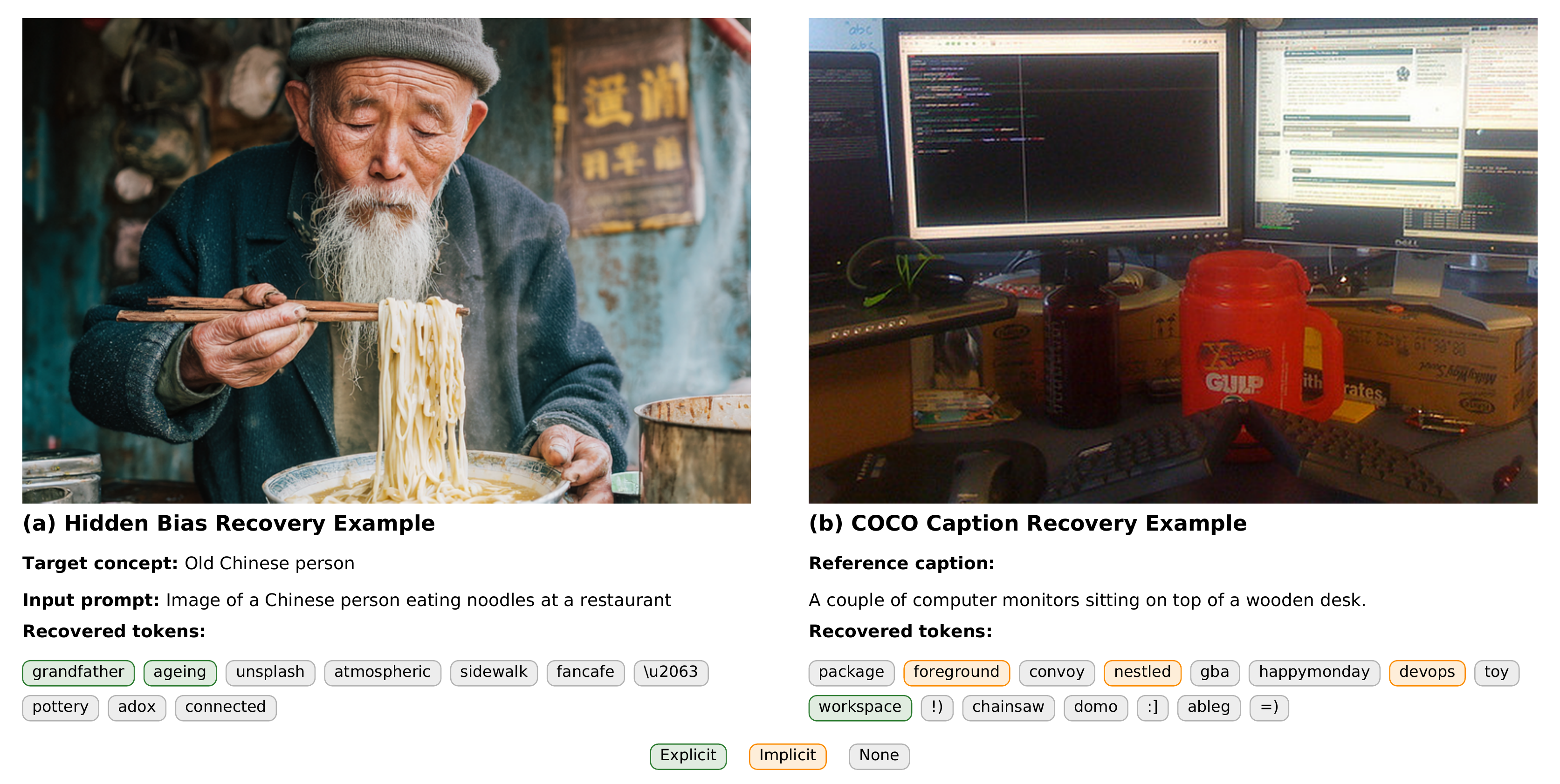}
	\caption{Recovered sequences with each token colored by its grounding label (explicit, implicit, or unrelated; ``None'' in the legend denotes unrelated). Left: hidden concept recovery. Right: caption recovery on COCO. The token \texttt{\textbackslash u2063} is the invisible-separator character U+2063 in the CLIP vocabulary.}
	\label{fig:token_grounding}
	\vspace{-.1in}
\end{figure*}

\subsection{Embedding-Space Mutation}

EmbGA mutates each offspring position independently with probability $m$. For a selected position, EmbGA samples a random embedding $\hat{r}$ from the embedding table and moves the offspring embedding toward it,
\begin{equation}
\label{eq:mutation}
\hat{o}_{ji}^{\,\prime} = \alpha \hat{o}_{ji} + (1-\alpha)\hat{r}, \qquad \alpha \sim U(0,1).
\end{equation}
Discrete mutation replaces a token with a random token, so the content of that position is lost and the size of the change depends only on the vocabulary. In contrast, Eq.~\eqref{eq:mutation} keeps a fraction $\alpha$ of the current embedding. The change is therefore bounded, and the mutated position stays related to its original content. As a result, the mutation rate controls how often the search changes a candidate rather than how much of the candidate is lost. A higher mutation rate can then be used for exploration without destroying good candidates. Section~\ref{sec:ablation} shows that this operator is the main reason EmbGA remains stable at high mutation rates.

\subsection{Projection Back to the Token Space}

Crossover and mutation produce embeddings that usually do not match any entry of the embedding table. EmbGA therefore projects each offspring position onto the vocabulary before scoring the candidate,
\begin{equation}
\label{eq:projection}
t^{*} = \arg\min_{t\in\mathcal{V}} \left\| e - E(t) \right\|_{2},
\end{equation}
where $e$ is the offspring embedding at that position. Eq.~\eqref{eq:projection} keeps EmbGA within the auditing model of Section~\ref{sec:problem}, since scoring needs only a forward pass and the returned result is a readable string. The projection also acts as an implicit regularizer, because it maps embeddings in regions with no nearby token back to the region that the encoder was trained on.

\subsection{Population Update}

EmbGA scores the projected offspring with Eq.~\eqref{eq:fitness} and updates the population by deterministic crowding. Each offspring competes with its most similar parent and replaces that parent only if the offspring has a higher score. Since each offspring competes only with a similar parent, the population keeps multiple distinct lineages, and no single lineage can take over the population.

\section{Experiments}

\subsection{Experimental Setup}

\begin{figure*}[t]
	\centering
	\includegraphics[width=\textwidth]{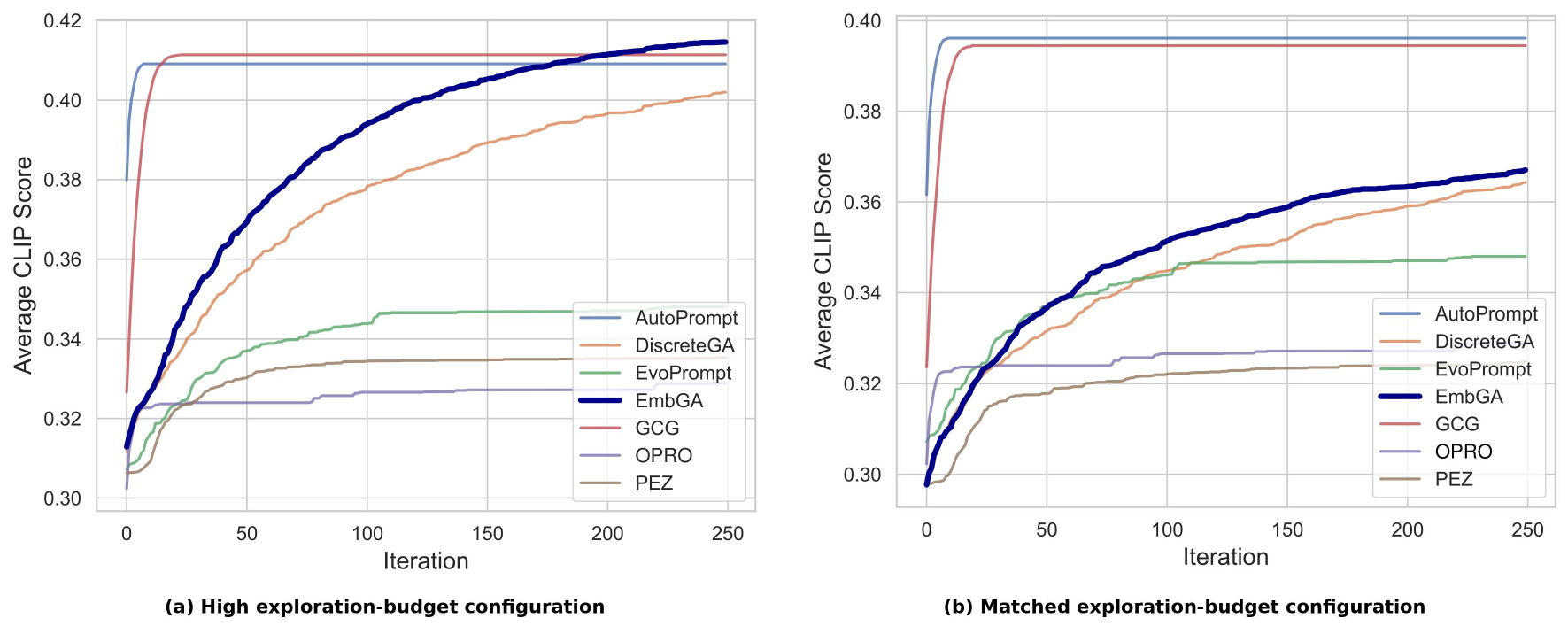}
	\caption{Alignment achieved on the hidden concept benchmark, averaged over the six categories, as a function of optimization iteration. Panel (a) uses the high exploration budget and panel (b) equalizes the budget across all methods.}
	\label{fig:clip_curves}
\end{figure*}

\myparatight{Benchmarks.} We evaluate EmbGA on two benchmarks whose target semantics come from different sources. The first is a hidden concept recovery benchmark based on Backdooring Bias~\cite{naseh2025backdooringbiasb2stable}. Table~\ref{tab:cases} lists its six cases, each of which contains a clean prompt, a hidden concept absent from the prompt, and images that contain the concept. For each case, we use a large language model to combine the clean prompt and the hidden concept into ten candidate prompts, and then use Midjourney to generate fifty images for each candidate prompt. This process gives five hundred images per category. The hidden concept is used only for evaluation, so the task tests recovery rather than verification. The second benchmark uses the first three hundred images of the COCO val2017 split~\cite{lin2015microsoftcococommonobjects}. Its target semantics come from natural images, so this benchmark tests whether EmbGA generalizes beyond the controlled hidden concept setting.

\begin{table*}[t]
	\centering
	\small
	\setlength{\tabcolsep}{5pt}
	\begin{tabular}{llccccccc}
		\toprule
		\textbf{Budget} & \textbf{Category}
		& \textbf{AutoPrompt}
		& \textbf{DiscreteGA}
		& \textbf{EmbGA}
		& \textbf{EvoPrompt}
		& \textbf{GCG}
		& \textbf{OPRO}
		& \textbf{PEZ}
		\\
		\midrule
		\multirow{7}{*}{High}
		& Age & $0.143$ & $0.145$ & $\mathbf{0.209}$ & $0.182$ & $0.115$ & $0.091$ & $0.000$ \\
		& Gender & $0.088$ & $0.024$ & $0.048$ & $0.091$ & $0.055$ & $\mathbf{0.136}$ & $0.069$ \\
		& Item & $0.055$ & $0.064$ & $0.065$ & $\mathbf{0.091}$ & $0.037$ & $\mathbf{0.091}$ & $0.000$ \\
		& Political Obj. & $0.028$ & $0.018$ & $0.038$ & $0.000$ & $0.057$ & $\mathbf{0.091}$ & $0.000$ \\
		& Political Sur. & $0.095$ & $0.093$ & $\mathbf{0.120}$ & $0.000$ & $0.113$ & $0.000$ & $0.046$ \\
		& Race & $0.073$ & $0.009$ & $0.045$ & $\mathbf{0.100}$ & $0.056$ & $0.050$ & $0.027$ \\
		& Average & $0.080$ & $0.059$ & $\mathbf{0.088}$ & $0.077$ & $0.072$ & $0.077$ & $0.024$ \\
		\midrule
		\multirow{7}{*}{Matched}
		& Age & $0.056$ & $0.118$ & $0.120$ & $\mathbf{0.182}$ & $0.102$ & $0.091$ & $0.038$ \\
		& Gender & $0.027$ & $0.018$ & $0.036$ & $0.091$ & $0.076$ & $\mathbf{0.136}$ & $0.000$ \\
		& Item & $0.000$ & $0.018$ & $0.009$ & $\mathbf{0.091}$ & $0.068$ & $\mathbf{0.091}$ & $0.009$ \\
		& Political Obj. & $0.018$ & $0.000$ & $0.000$ & $0.000$ & $0.000$ & $\mathbf{0.091}$ & $0.000$ \\
		& Political Sur. & $0.049$ & $0.038$ & $0.009$ & $0.000$ & $\mathbf{0.073}$ & $0.000$ & $0.027$ \\
		& Race & $0.060$ & $0.018$ & $0.027$ & $\mathbf{0.100}$ & $0.018$ & $0.050$ & $0.000$ \\
		& Average & $0.035$ & $0.035$ & $0.034$ & $\mathbf{0.077}$ & $0.056$ & $\mathbf{0.077}$ & $0.012$ \\
		\bottomrule
	\end{tabular}
	\caption{Token grounding ($\uparrow$) per category under the high and matched exploration budgets. Under the high budget, EmbGA achieves the best average, and its margin over DiscreteGA shows the gain from changing the operator space at the same population size. Under the matched budget, the language-model-guided optimizers achieve the best grounding because of their semantic prior.}
	\label{tab:grounding_high}
	\vspace{-.1in}
\end{table*}

\myparatight{Compared methods.} We compare EmbGA with six methods from the three families in Section~\ref{sec:related}. The gradient-based family includes GCG~\cite{zou2023universal}, AutoPrompt~\cite{shin2020autopromptelicitingknowledgelanguage}, and PEZ~\cite{wen2023hardpromptseasygradientbased}. We give these methods backward access to the scoring encoder, which EmbGA does not use, so their results serve as an upper reference rather than a direct comparison. The evolutionary family is represented by DiscreteGA, a discrete-token genetic baseline based on GAAPO~\cite{secheresse2025gaapogeneticalgorithmicapplied}. DiscreteGA keeps the midpoint crossover of GAAPO and replaces selected positions with uniformly sampled vocabulary tokens during mutation. It serves as a direct ablation of EmbGA, since the two methods differ only in the space where the operators act. The language-model-guided family includes EvoPrompt~\cite{guo2025evopromptconnectingllmsevolutionary} and OPRO~\cite{yang2024largelanguagemodelsoptimizers}, both of which use Qwen2.5-14B-Instruct in BF16~\cite{qwen2025qwen25technicalreport}.

\myparatight{Evaluation metrics.} We report two complementary metrics. The first is the CLIP alignment score, which measures how well the recovered prompt matches the target images. Given normalized image embeddings $\{v_i\}_{i=1}^{N}$ and the normalized text embedding $u$ of the recovered prompt, the score is defined as
$
\frac{1}{N}\sum_{i=1}^{N}\max\!\left(\cos(v_i,u),0\right).
$
Unlike the fitness in Eq.~\eqref{eq:fitness}, this metric clips negative similarities to zero. Unlike CLIPScore~\cite{hessel2022clipscorereferencefreeevaluationmetric}, it uses no calibration, because our goal is to compare optimizers rather than to match human judgment. However, a sequence can obtain a high alignment score by exploiting weaknesses of the encoder while containing no words that a human would recognize. We therefore also report a token-level grounding score (TLGS). A separate language model labels each recovered token as explicit, implicit, or unrelated with respect to the target concept or reference caption, and these labels receive scores of $1.0$, $0.5$, and $0.0$, respectively. For a recovered sequence of $M$ tokens, the score is
$
\mathrm{TLGS} = \frac{1}{M}\sum_{i=1}^{M}s(t_i)$ with $s(t_i)\in\{0,0.5,1\}$. We use Qwen3-32B in BF16~\cite{qwen2025qwen25technicalreport} as the evaluator. Since TLGS relies on model judgment, we use it to compare methods rather than as an absolute measure of correctness. Figure~\ref{fig:token_grounding} shows examples of these labels.

\myparatight{Parameter setup.} All methods use the same CLIP ViT-B/32 encoder for scoring, append the recovered tokens to the clean prompt after a comma, and use a fixed sequence length of ten tokens for hidden concept recovery and fifteen tokens for caption recovery. EmbGA and DiscreteGA both use a tournament size of three and a mutation rate of $0.1$. We run each stochastic method five times with different random seeds and report the mean and standard deviation. Because of their high cost, the language-model-guided baselines are run once. Since Challenge 3 links exploration to compute, we evaluate the hidden concept benchmark under two budget settings. In the high-budget setting, EmbGA and DiscreteGA use a population of $4000$, GCG and AutoPrompt use a top-$k$ of $512$, and PEZ optimizes a batch of $1000$ initializations and returns the best one. EvoPrompt and OPRO use $100$ candidates, which is the largest number their per-candidate cost allows. In the matched-budget setting, every method is limited to $100$ candidates, top-$k$ values, or batch entries, which separates the effect of algorithm design from the effect of scale.

\subsection{Main Results}

\myparatight{EmbGA achieves the highest alignment under the high budget.} Figure~\ref{fig:clip_curves}(a) shows the alignment averaged over the six categories under the high budget. EmbGA achieves the highest average score among all seven methods, even though it does not use the gradients that GCG, AutoPrompt, and PEZ use. The most direct comparison is with DiscreteGA, which uses the same population size, selection rule, replacement rule, and mutation rate as EmbGA. Since the two methods differ only in whether the operators act on embeddings or on token indices, the advantage of EmbGA comes from the operator space. This result supports the continuity property in Section~\ref{sec:principles}.

\myparatight{Gradient-based methods lead under the matched budget.} Figure~\ref{fig:clip_curves}(b) limits every method to one hundred candidates. In this setting, AutoPrompt and GCG achieve the highest alignment. This result is expected, because a gradient points to a promising substitution without evaluating it, which is most useful when only a few evaluations are allowed. EmbGA remains the best evolutionary method and still outperforms DiscreteGA by a small margin. The two panels thus show a trade-off between gradient information and exploration. Gradient-based methods perform best when exploration is limited, while EmbGA uses additional evaluations more effectively and takes the lead as the budget grows. When backward access to the scoring encoder is unavailable but forward evaluations are affordable, the high-budget setting is the relevant one.

\myparatight{Higher alignment does not always lead to better grounding.} Table~\ref{tab:grounding_high} reports token grounding under the two budgets. Under the matched budget, EvoPrompt and OPRO achieve the highest average grounding of $0.077$, compared with $0.034$ for EmbGA and $0.035$ for DiscreteGA, despite their lower alignment. A language model tends to propose words related to the target domain, whereas a search without such guidance can reach high alignment with tokens that the evaluator labels as unrelated. Under the high budget, this advantage disappears. EmbGA achieves the highest average grounding of $0.088$, compared with $0.077$ for both language-model-guided methods and $0.059$ for DiscreteGA. With enough exploration, EmbGA thus finds semantically related tokens without a language-model prior. The per-category results show that no method is best in every category. EmbGA leads on Age and Political Surround, while OPRO leads on Gender and Political Object. We attribute this variation to differences in concept size and to the fact that one sequence of ten tokens must describe an entire image set. In addition, the absolute grounding scores are low for all methods, which shows that recovering a complete semantic description remains difficult. These results therefore show relative improvements in token grounding rather than reliable recovery of the full hidden concept.

\begin{figure}[t]
	\centering
	\includegraphics[width=\columnwidth]{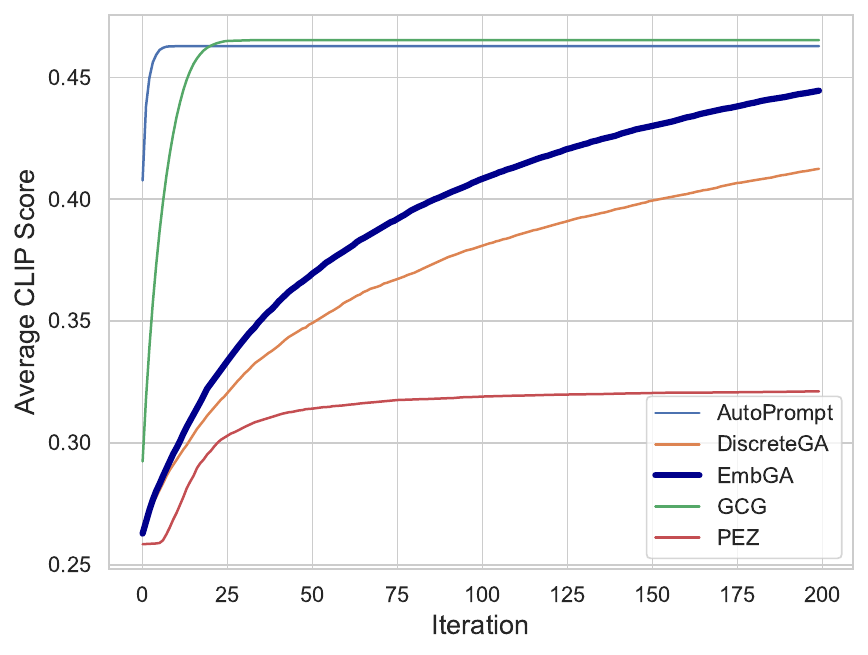}
	\caption{Alignment over the course of optimization on COCO caption recovery, where every method is given the same compute budget rather than the same number of candidates.}
	\label{fig:coco_curve}
	\vspace{-.15in}
\end{figure}

\begin{figure*}[t]
	\centering
	\includegraphics[width=0.8\textwidth]{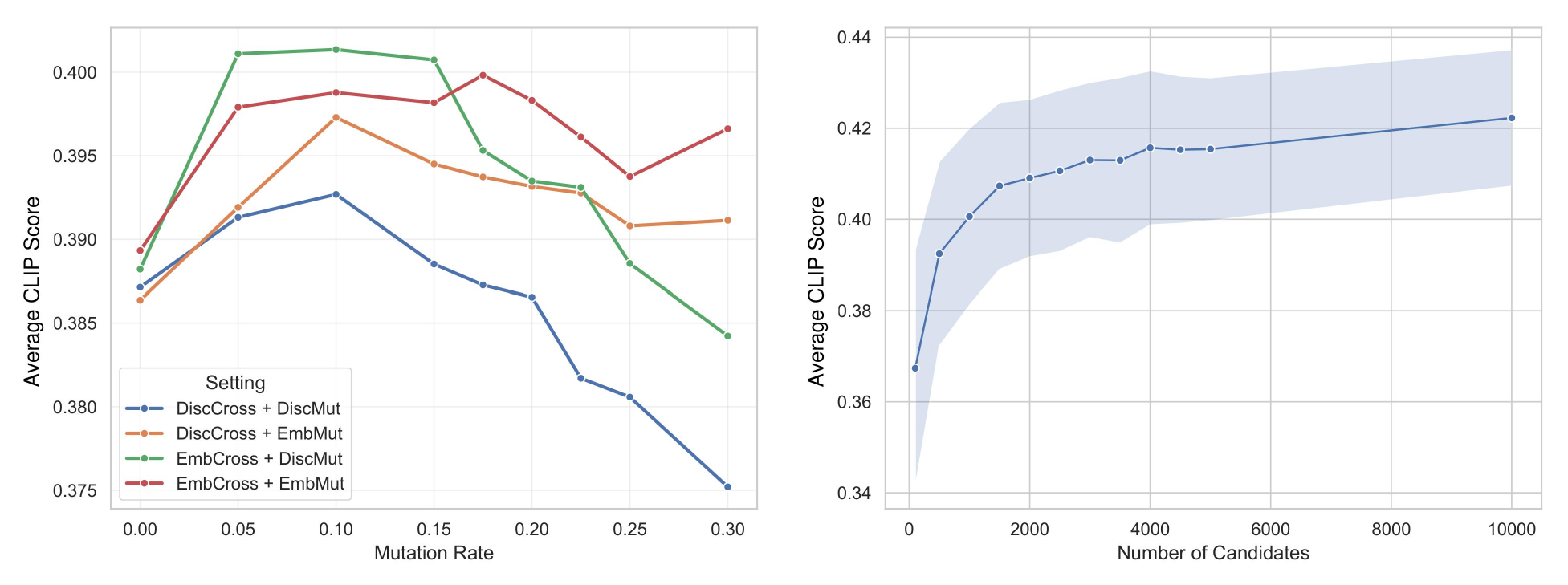}
	\caption{Ablation study of EmbGA. Left: alignment of the four combinations of discrete and embedding-space crossover and mutation under different mutation rates, with a population of one thousand. Right: alignment under different population sizes for the full embedding-space configuration with a mutation rate of $0.1$. Each point is averaged over five runs.}
	\label{fig:operator_ablation}
	\vspace{-.1in}
\end{figure*}

\begin{table}[t]
\centering
\setlength{\tabcolsep}{5pt}
\begin{tabular}{lccc}
\toprule
\textbf{Method}
& \textbf{CLIP Score}
& \textbf{TLGS}
& \textbf{Runtime (s)}
\\
\midrule
AutoPrompt
& $0.463 \pm 0.028$
& $0.095$
& $201.69 \pm 2.03$
\\
DiscreteGA
& $0.413 \pm 0.021$
& $0.109$
& $205.57 \pm 1.27$
\\
EmbGA
& $0.445 \pm 0.020$
& $\mathbf{0.114}$
& $194.19 \pm 1.26$
\\
GCG
& $\mathbf{0.465 \pm 0.028}$
& $0.089$
& $190.81 \pm 0.95$
\\
PEZ
& $0.321 \pm 0.016$
& $0.067$
& $207.69 \pm 1.29$
\\
\bottomrule
\end{tabular}
\caption{Caption recovery on COCO under a fixed compute budget of roughly two hundred seconds. EmbGA recovers the most grounded tokens, while GCG and AutoPrompt achieve slightly higher alignment.}
\label{tab:coco_main}
% \vspace{-.1in}
\end{table}

\myparatight{EmbGA also recovers ordinary semantics on COCO.} Table~\ref{tab:coco_main} reports caption recovery on COCO, and Figure~\ref{fig:coco_curve} shows the optimization curves. Since the per-iteration cost differs greatly across method families, we fix a compute budget of about two hundred seconds and configure each method to use this budget: a population of $2200$ for DiscreteGA and $2000$ for EmbGA, a top-$k$ of $256$ for AutoPrompt and $276$ for GCG, and a batch of $1150$ initializations for PEZ. EvoPrompt and OPRO are excluded because their runtime far exceeds the budget. GCG and AutoPrompt achieve the highest alignment of $0.465$ and $0.463$, respectively, and EmbGA follows at $0.445$ without using gradients. The ranking is different for grounding, where EmbGA achieves the best score of $0.114$, followed by DiscreteGA at $0.109$, AutoPrompt at $0.095$, GCG at $0.089$, and PEZ at $0.067$. This result, together with the examples in Figure~\ref{fig:token_grounding}, indicates that the tokens recovered by EmbGA more often describe the image content instead of only matching the encoder, which is important for auditing because the output of an audit is read by a human.

\begin{table}[t]
\centering
\small
\setlength{\tabcolsep}{5pt}
\begin{tabular}{lcc}
\toprule
\textbf{Method} & \textbf{High} & \textbf{Matched} \\
\midrule
AutoPrompt & $312.68 \pm 2.00$ & $78.26 \pm 0.04$ \\
DiscreteGA & $461.30 \pm 2.27$ & $\mathbf{12.88 \pm 0.02}$ \\
EmbGA & $473.14 \pm 2.36$ & $13.19 \pm 0.03$ \\
EvoPrompt & $15318.61$ & $15318.61$ \\
GCG & $293.63 \pm 1.43$ & $59.77 \pm 0.03$ \\
OPRO & $26868.00$ & $26868.00$ \\
PEZ & $\mathbf{223.24 \pm 1.32}$ & $23.67 \pm 0.06$ \\
\bottomrule
\end{tabular}
\caption{Wall-clock time in seconds ($\downarrow$) of one recovery run on the hidden concept benchmark under the two budgets. Entries without standard deviations are single runs of the language-model-guided methods.}
\label{tab:runtime}
 \vspace{-.1in}
\end{table}

\myparatight{EmbGA adds little cost over DiscreteGA and is much cheaper than language-model-guided search.} Table~\ref{tab:runtime} reports the wall-clock time of one recovery run on the hidden concept benchmark. Under the high budget, EmbGA takes $473.14$ seconds and DiscreteGA takes $461.30$ seconds, so interpolation and nearest-token projection add only about $2.5\%$ to the runtime. In contrast, EvoPrompt takes $15318.61$ seconds and OPRO takes $26868.00$ seconds, which are about $32$ and $57$ times the runtime of EmbGA, even though they evaluate only one fortieth as many candidates. Their runtimes stay the same under both budgets, because they already use $100$ candidates in the high-budget setting and cannot be scaled further. Under the matched budget, EmbGA takes $13.19$ seconds, which is close to DiscreteGA at $12.88$ seconds and lower than PEZ at $23.67$, GCG at $59.77$, and AutoPrompt at $78.26$ seconds, because each gradient step costs more than a forward pass.

\subsection{Ablation Studies}
\label{sec:ablation}

Figure~\ref{fig:operator_ablation} separates the effects of crossover and mutation. The fully discrete configuration performs worst at every nonzero mutation rate, and replacing either operator with its embedding-space version improves alignment. Embedding-space crossover helps mainly at low mutation rates. Embedding-space mutation keeps performance stable as the mutation rate increases, which shows that bounded mutation is the main source of robustness. Performance also improves with population size and begins to saturate between $2000$ and $4000$ candidates. These results show that the gains of EmbGA come from its operators rather than from population size alone.

\section{Limitations and Conclusion}

We propose EmbGA, which moves genetic crossover and mutation from discrete token indices to the embedding space and projects every candidate back to vocabulary tokens before evaluation. On two benchmarks, EmbGA outperforms discrete evolutionary search, achieves the highest alignment when the exploration budget is large, and achieves the highest grounding score on COCO, while running much faster than language-model-guided methods. Gradient-based methods remain a better choice when backward access is available and the evaluation budget is small.

Our evaluation covers only one hidden concept setup and a subset of COCO, so its generalization to other encoders, generators, and concept types remains open. In addition, EmbGA requires the token embedding table of the auxiliary CLIP model, and TLGS relies on a language-model evaluator. Since a high alignment score does not guarantee that the recovered tokens describe concepts absent from the prompt, the recovered sequences should be treated as candidate descriptions rather than definitive identifications.

\bibliography{Bibliography-File}

\end{document}